\documentstyle[preprint,aps,epsf,epsfig]{revtex}

\begin{document}
\draft
\title{
Magnetic skyrmions and their lattices in triplet superconductors}
\author{A. Knigavko$^1$, B. Rosenstein$^{1,2}$, and Y. F. Chen$^1$}
\address{
$^1$Electrophysics Department, National Chiao Tung University\\
Hsinchu, Taiwan 30043, Republic of China\\
$^2$National Center for Theoretical Sciences, National Chiao Tung University\\
Hsinchu, Taiwan 30043, Republic of China\\
}

\maketitle
\centerline{\small (Last revised January 23, 1999)}

\begin{abstract}
Complete topological classification of solutions in $SO(3)$ symmetric
Ginzburg--Landau free energy has been performed and a new class of solutions
in weak external magnetic field carrying two units of magnetic flux has been
identified. These solutions, magnetic skyrmions, do not have singular core
like Abrikosov vortices and at low magnetic field become lighter for
strongly type II superconductors. As a consequence, the lower critical
magnetic field $H_{c1}$ is reduced by a factor of $\log \kappa $. Magnetic
skyrmions repel each other as $1/r$ at distances much larger then magnetic
penetration depth $\lambda $ forming relatively robust triangular lattice.
Magnetization near $H_{c1}$ increases gradually as $(H-H_{c1})^{2}$. This
behavior agrees very well with experiments on heavy fermion superconductor $%
UPt_{3}$. Newly discovered $Ru$ based compounds $Sr_{2}RuO_{4}$ and $%
Sr_{2}YRu_{1-x}Cu_{x}O_{6}$ are other possible candidates to possess
skyrmion lattices. Deviations from exact $SO(3)$ symmetry are also studied.
\end{abstract}

\pacs{PACS numbers: 74.20.De, 74.25.Ha, 74.60.Ec, 74.70.Tx}

%


\section{Introduction}

Rich variety of novel magnetic properties can be found in superconductors
with unconventional type of pairing symmetry. At present several examples of
unconventional superconductors are known. The first of them is, of course, a
family of $T_{c}$ cuprates. In connection with them a lot of recent efforts
were devoted to study of $d$ - wave pairing of various types with possible
admixture of $s$ - wave. On the other hand, a triplet type of pairing is
believed to exist in $UPt_{3}$\cite{Machida1,Sauls1} and in some other heavy
fermion compounds\cite{Sigrist1}. It is also suspected to occur in recently
discovered new classes of $Ru$ based superconductors, layered perovskite $%
Sr_{2}RuO_{4\text{ }}$\cite{Maeno,Rice} and bulk compound $%
Sr_{2}YRu_{1-x}Cu_{x}O_{6}$\cite{Wu} which has a double perovskite
structure. Since all mentioned superconductors are of strongly type II,
vortices play the major role in their thermodynamical properties. In high $%
T_{c}$ superconductors, despite fundamental differences in mechanism and
microscopic properties compared to conventional superconductors, vortices
are quite similar to conventional Abrikosov vortices. The reason is that
there exists a dominant single order parameter field: $d$ - wave condensate.
Small (sometimes quite important) deviations can be accounted for due to
admixture of the $s$ - wave component. Then, the order parameter is
effectively multicomponent. This property leads generally to various new
effects like nonaxisymmetric vortices\cite{Ting,Berlinsky} and phase
transitions within flux line lattices near $H_{c2}$\cite{Rosenstein}.
Similar phenomena exist and should be even more pronounced in the systems
with intrinsically multicomponent superconducting order parameter\cite
{Tokuyasu,Luk'yanchuk} like heavy fermions compounds.

The situation in triplet superconductors might be more exotic. The order
parameter is necessarily multicomponent. In addition, under certain
conditions the rotational symmetry (at least approximate one) between
different components might exist. In that case vortices are not the only
type of topological solitons which can carry magnetic flux through the
sample. The corresponding phenomenological Ginzburg-Landau theory has the
order parameter of a vector type with a continuous symmetry. It is known in
the theory of superfluid $^{3}He$ \cite{Salomaa} that in such a system there
exist topological defects which have no singularities even within the London
approximation. On the other hand, vortices have a singularity at their core,
at least for one of components of the order parameter. This makes their
energy roughly proportional to $\log \kappa ,$ similarly to the case of a
standard Abrikosov vortex. Therefore, for sufficiently large $\kappa $
vortices are expected to be heavier then nonsingular topological defects and
the latter become most likely candidates for thermodynamically stable
configurations of the order parameter field into which a homogeneous
superconducting (Meissner) state transforms under the action of an external
magnetic field.

It is the purpose of the present paper to investigate this possibility in
detail. We find such a nonsingular in the London approximation solution, the
magnetic skyrmion, describe its structure and show that it is energetically
favorable over Abrikosov vortex in wide range of Ginzburg-Landau parameter $%
\kappa $ values. Lattices of magnetic skyrmions are particularly important
at fields near the lower critical field. Most striking effects are reduction
of $H_{c1}$ by a factor of $\log \kappa $ and dramatic change in the
behavior of magnetization near $H_{c1}$. We also investigate what happens to
magnetic skyrmions when the continuous symmetry breaking terms are
introduced into the free energy. It is shown they survive under small
perturbations and gradually evolve to another still nonsingular
configurations under large perturbations of a certain type.

Magnetic skyrmion lattices may have been already experimentally observed in $%
UPt_{3}$. Magnetization curves of \ near $H_{c1}$ \cite{Maple,Zhao} are
rather unusual (see Fig.1 of \cite{Knigavko1} in which a short account of
this work was presented). Theoretically, if the magnetization is due to
penetration of vortices into a superconducting sample then one expects $%
-4\pi M$ to drop with an infinite derivative at $H_{c1}$. On the other hand
experimentally $-4\pi M$ continues to increase smoothly. Such a behavior was
attributed to strong flux pinning or surface effects \cite{Maple}. However
both experimental curves in Fig.1 of \cite{Knigavko1}, as well as the other
ones found in literature, are close to each other if plotted in units of $%
H_{c1}$. We propose a more fundamental explanation of the universal smooth
magnetization curve near $H_{c1}$. If one assumes that fluxons are of
unconventional type for which interaction is long range \ then precisely
this type of magnetization curve is obtained. Indeed magnetization near $%
H_{c1}$ due to fluxons carrying $N$ units of flux $\Phi _{0}\equiv hc/2e$,
with line energy $\varepsilon $ and mutual interaction $V(r)$, is found by
minimizing the Gibbs energy of a very sparse triangular lattice: 
\begin{equation}
G(B)=\frac{B}{N\Phi _{0}}\left[ \varepsilon +3V(a_{\bigtriangleup })\right] -%
\frac{BH}{4\pi },  \label{Gibbs-energy}
\end{equation}
where $a_{\bigtriangleup }=(\Phi _{0}/B\sqrt{3})^{\frac{1}{2}}$ is lattice
spacing. When $V(r)\sim \exp [-\lambda r],$ the magnetic induction has the
conventional behavior $B\sim \left[ \log \left( H-H_{c1}\right) \right]
^{-2} $\cite{Tinkham}, while if it is long range, $V(r)\sim 1/r^{n},$ then
one finds $B\sim \left( H-H_{c1}\right) ^{n+1}$. The physical reason for
this different behavior is very clear. For a short range repulsion, if one
fluxon penetrated the sample, many more can penetrate almost with no
additional cost of energy. This leads to the infinite derivative of
magnetization. On the other hand for a long range interaction making a place
for each additional fluxon becomes energy consuming. Derivative of
magnetization thus becomes finite.

The remainder of the paper is organized as follows. In section II we present
the $SO(3)$ symmetric model and note that it is an excellent approximation
to certain successful models of $UPt_{3}$\cite{Machida2} as well as to
others \cite{Zhitom1}. The London approximation is developed. In section III
we perform complete topological classification of solutions and find that
the magnetic skyrmion carries two units of magnetic flux. General form of
cylindrically symmetrical solutions is given. In section IV we determine
magnetic skyrmion lattice structure, $H_{c1}$ and the magnetization curve.
An example of deviations from exact $SO(3)$ symmetry is considered in
section V. More specifically, we address the case of Zeeman like interaction
relevant to $Sr_{2}YRu_{1-x}Cu_{x}O_{6}$\cite{Knigavko2} system which
initially motivated us to search for exotic vortices. Section VI contains
discussion of the results and possibilities to experimentally observe
various effects of magnetic skyrmions.

\section{The model}

\subsection{Ginzburg-Landau free energy functional and its symmetries}

Let us consider a model Ginzburg--Landau theory with the order parameter $%
\psi _{i}(\vec{r})$ being a three dimensional ($i=1,2,3)$ complex vector. It
is convenient to consider index $\ i$ as a spin in the case of weak
spin-orbit coupling in the pairing channel, but this is not necessary
interpretation. The case of strong spin - orbit interaction can also be
addressed provided some modifications of the free energy functional are
made. Average spin of the Cooper pair at a specific point in the space is
given by 
\begin{equation}
S_{i}(\vec{r})\equiv \psi _{j}^{\ast }(\vec{r})\left( -i\varepsilon
_{i\,jk}\right) \psi _{k}(\vec{r}).  \label{spin}
\end{equation}
The material under study is assumed to be isotropic. Extensions of our
results to anisotropic situations are discussed in the section V. \ 

The Ginzburg - Landau free energy functional of the system has the form: 
\begin{eqnarray}
F &=&F_{pot}+F_{grad}+\frac{1}{8\pi }B_{j}^{2},  \label{Free energy} \\
F_{pot} &=&-\alpha \psi _{i}\psi _{i}^{\ast }+\frac{\beta _{1}}{2}(\psi
_{i}\psi _{i}^{\ast })^{2}+\frac{\beta _{2}}{2}|\psi _{i}\psi _{i}|^{2},
\label{Fpot} \\
F_{grad} &=&\frac{\hbar ^{2}}{2m^{\ast }}({\cal D}_{j}\psi _{i})({\cal D}%
_{j}\psi _{i})^{\ast },  \label{Fgrad}
\end{eqnarray}
where ${\cal D}_{j}\equiv \partial _{j}-i({e^{\ast }}/{\hbar c})A_{j}$ are
covariant derivatives, $B_{j}\equiv (\nabla \times \vec{A})_{j},$ $m^{\ast
}>0$ is effective mass of the pair and $e^{\ast }$ is effective charge of
the pair. For a superconducting phase to exist the coefficient $\alpha $
should be positive below the phase transition point and we set $\alpha
=\alpha ^{\prime }(T_{c}-T)$ with $\alpha ^{\prime }>0,$ while for positive
definiteness of the potential the other coefficients of eq.(\ref{Fpot})
should satisfy $\beta _{1}>0$ and $\beta _{2}>-\beta _{1}.$

The free energy density eq.(\ref{Free energy}) has the following independent
symmetries. The spin rotations, forming a group $SO_{spin}(3),$ act on the
index $i$ of the order parameter field, so that it transforms as a vector.
Two dimensional (orbital) space rotations, forming a different $%
SO_{orbit}(2) $ group, act on spatial coordinates $x_{j}$ and the electric
charge transformations, forming a $U(1)$ group, rotate the complex phase of
the order parameter. Note that in eq.(\ref{Free energy}) we assumed that
external magnetic field $\vec{H}$ is oriented along $\ z$ direction. We will
consider only configurations invariant under translations in that direction
or the thin film geometry.

First we consider the case of zero external magnetic field. Pure
superconducting (Meissner) phases that appear below $T_{c}$ are found by
minimization of $F_{pot}$ with respect to $\psi _{i}^{\ast }.$ This is
conveniently done making use of the following parametrization of the order
parameter vector: 
\begin{equation}
\vec{\psi}=\psi _{i}\vec{e}_{i}=f(\vec{n}\cos \phi +i\,\vec{m}\sin \phi ),
\label{parametrization}
\end{equation}
where $f>0$, $0\leq \phi \leq \pi /2$, $\vec{n}$ and $\vec{m}$ are unit
vectors that are arbitrarily oriented with respect to some fixed coordinate
system in the spin space with orthonormal basis $\vec{e}_{1},\vec{e}_{2},%
\vec{e}_{3}$. There exist two phases, depending on the sign of the
coefficient $\beta _{2}:$

\begin{eqnarray}
\text{I} &:&\text{ }\beta _{2}>0\;\hspace{\stretch{2}}\hspace{1cm}\vec{\psi}%
=f\frac{\vec{n}+i\vec{m}}{\sqrt{2}},\quad \;\vec{n}\perp \vec{m},\,\phi =\pi
/4,\quad f^{2}=\frac{\alpha }{\beta _{1}}.  \label{phaseI} \\
\text{II} &:&\text{{}}\beta _{2}<0\;\hspace{\stretch{2}}\hspace{1cm}\;%
\hspace{\stretch{2}}\vec{\psi}=fe^{i\phi }\vec{n},\qquad \vec{n}=\pm \vec{m}%
;\quad f^{2}=\frac{\alpha }{\beta _{1}+\beta _{2}}.  \label{phaseII}
\end{eqnarray}
Combining eq.(\ref{spin}) and eq.(\ref{parametrization}) one obtains $\vec{S}%
=f^{2}\sin {2\phi }\;\vec{l},$ where $\vec{l}\equiv \vec{n}\times \vec{m}$.
In the phase I the projection of the spin of a Cooper pair $\vec{S}$ on the
vector $\vec{l}$ is equal to either $+1$ or $-1$, reflecting spontaneous
time reversal symmetry breaking. In the phase II this projection is always
zero.

\subsection{Theories of triplet superconductors and terms breaking $%
SO_{spin}(3)$ symmetry}

Obviously, the model of the previous subsection is an idealization of the
actual situation in triplet superconductors. In this subsection we note that
some successful models of $UPt_{3},$ notably that of Machida {\it et al} 
\cite{Machida2}, differ from this model only by small less symmetric terms.
These terms could be considered as small perturbations, at least in some
regions of $H-T$ diagram. The asymmetries are of several types. First, the
space symmetry $SO_{orbit}(2)$ is normally broken down to some
crystallographic point group of a given material ($D_{6h}$ for $UPt_{3}$, $%
D_{4h}$ for $Sr_{2}RuO_{4}$, $D_{2h}$ for $Sr_{2}YRu_{1-x}Cu_{x}O_{6}$). The
effective mass $m^{\ast }$ then becomes a symmetric tensor $m_{jk}^{\ast }$.
Second, the spin $\vec{S}$ can be coupled to the magnetic field. This
explicitly breaks $SO_{spin}(3)$ down to $SO_{spin}(2)$. Separate spin and
orbital symmetries $SO_{spin}(2)\otimes SO_{orbit}(2)$ are broken down to
diagonal $SO_{tot}(2)$ as well. The various types of perturbations are 
\begin{mathletters}
\begin{eqnarray}
\Delta F_{pot2} &=&\alpha ^{\prime }\left[ (T_{c}-T_{c}^{^{(1)}})|\psi
_{1}|^{2}+(T_{c}-T_{c}^{^{(2)}})|\psi _{2}|^{2}\right] ,  \label{deva} \\
\Delta F_{pot4} &=&\beta _{3}(|\psi _{3}|^{2}-|\psi _{1}|^{2}-|\psi
_{2}|^{2})^{2}+\beta _{4}|\psi _{3}|^{2}(|\psi _{3}|^{2}-|\psi
_{1}|^{2}-|\psi _{2}|^{2}),  \label{devb} \\
\Delta F_{grad} &=&K\left[ ({\cal D}_{j}\psi _{i})({\cal D}_{i}\psi
_{j})^{\ast }+({\cal D}_{i}\psi _{i})({\cal D}_{j}\psi _{j})^{\ast }\right] ,
\label{devc} \\
\Delta F_{Zeeman} &=&\mu \vec{S}\cdot \vec{B},  \label{devd} \\
\Delta F_{nonlin} &=&\Delta \chi |\vec{\psi}\cdot \vec{B}|^{2}.  \label{deve}
\end{eqnarray}

We estimate their coefficients for the case of $UPt_{3}$. Some models of $%
UPt_{3}$ do not have three dimensional complex order parameter and therefore
will not be addressed here. Examples include $E_{1g}$ singlet pairing \cite
{Joynt}, $E_{2g}$triplet pairing \cite{Sauls2}, accidental degenerate $AB$
model \cite{Garg}. On the other hand, the model \ of weak spin--orbit
coupling developed by Machida {\it et al} \cite{Machida2} is of the type we
are interested in. In this model asymmetric terms are very important in
explaining the double superconducting phase transition at zero external
magnetic field. However, they are small in the low temperature
superconducting phase (phase B) well below its critical temperature $T\ll
T_{c}^{-}\simeq .45K$ and at low magnetic fields $H\simeq H_{c1}$. Indeed,
for the quadratic terms one gets from experiment $\frac{T_{c}-T_{c1}}{T_{c}}%
\sim .2$, $\frac{T_{c1}-T_{c2}}{T_{c}}<.05$. The quartic terms eq.(\ref{devb}%
) are order of magnitude smaller. The corrections to the gradient terms are
very small $K/\left( \frac{\hbar ^{2}}{2m^{\ast }}\right) \sim .01$ and $\mu
/$ $\left( \frac{\hbar ^{2}}{2m^{\ast }}\right) \sim .01$. The coefficient
of the nonlinear coupling term eq.(\ref{devc}) is negligible: $\left( \frac{%
\Delta \chi }{2}H_{c1}^{2}\right) /\left( \frac{\alpha ^{2}}{2\beta _{1}}%
\right) \simeq 10^{-6}.$ \ Another model of $UPt_{3},$ which has similar
structure to eq.(\ref{Free energy}), is the accidental degenerate $AE$ \cite
{Zhitom1}. Estimates are similar with exception of $K/\left( \frac{\hbar ^{2}%
}{2m^{\ast }}\right) $ which is now of order one (found from fitting
transition lines near $H_{c2}$).

In general, topological solitons exist even in those cases when the symmetry
is weakly broken. In section V we consider in detail influence of one
symmetry breaking term, Zeeman coupling, eq.(\ref{devb}), in connection to a
new material $Sr_{2}YRu_{1-x}Cu_{x}O_{6}$. We show that it do not affect
stability of solitons that we investigate in this paper

\subsection{London approximation}

The London approximation assumes that the order parameter has the form
determined by the potential part of the free energy eq.(\ref{Fpot}). In
particular, the modulus of the order parameter is fixed. Any variations of
the order parameter field over the space are only due to changes of the
degeneracy parameters which parametrize the vacuum manifold. From this
standpoint in usual s-wave superconductors there are no topological solitons
within the London approximation. Famous Abrikosov vortex has a core --a
region in which the modulus of the order parameter varies significantly and
vanishes at some point. A vortex can be incorporated into the London
approximation at the cost of singularities: vortex core is assumed to shrink
to a point in which energy diverges logarithmically. Accordingly, a cutoff,
the correlation length, should be introduced and one obtains $\log \kappa $
dependence for a vortex line tension. As discussed above, this means that if
there exists a nonsingular solution it is bound to become energetically
favorable for $\kappa $ large enough.

Below we concentrate on the properties of nonunitary phase I, eq.(\ref
{phaseI}), of a triplet superconductor near the lower critical field $%
H_{c1}. $ This phase is always assumed when we refer to the superconducting
state. We define magnetic penetration depth $\lambda \equiv \frac{c}{\mid
e^{\ast }\mid }\sqrt{\frac{\beta _{1}m^{\ast }}{4\pi \alpha }}$, coherence
length $\xi \equiv \hbar /\sqrt{2\alpha m^{\ast }}$, flux quantum $\Phi
_{0}\equiv hc/e^{\ast }$ and Ginzburg-Landau parameter $\kappa \equiv
\lambda /\xi .$ For convenience of the following discussion we express all
physical quantities in dimensionless units as follows:

\end{mathletters}
\begin{equation}
x\equiv \lambda \tilde{x},\;\;F\equiv \frac{\alpha ^{2}}{\beta _{1}\kappa
^{2}}\tilde{F},\quad f^{2}\equiv \frac{\alpha }{\beta _{1}}\tilde{f}%
^{2},\quad A\equiv \frac{\Phi _{0}}{2\pi \lambda }a\ ,\quad B\equiv \frac{%
\Phi _{0}}{2\pi \lambda ^{2}}\ b.\;\;  \label{units}
\end{equation}
The ''tilde'' marks will be omitted hereafter.

In order to determine the degeneracy parameters we consider the symmetry
breaking pattern of the superconducting state. Both the spin rotation $%
SO_{spin}(3)$ symmetry and the superconducting phase $U(1)$ symmetry are
spontaneously broken, but a diagonal subgroup $U(1)$ survives. It consists
of combined transformations: rotations by angle $\vartheta $ around the axis 
$\vec{l}$ which are accompanied by gauge transformations $e^{i\vartheta }$.
These combined transformations together with rotations of vector $\vec{l}$
itself, form the vacuum manifold. The vacuum manifold is isomorphic to the $%
SO(3)$ group. Our aim is to find nonsingular topological line defects in
this case. We choose a triad of orthonormal vectors $\vec{n},$ $\vec{m},$ $%
\vec{l}$ \ to be the degeneracy parameters. From the definition of these
vectors the following important relations can be derived: 
\begin{eqnarray}
\vec{n}\partial _{i}\vec{m} &=&-\vec{m}\partial _{i}\vec{n},  \label{rel1} \\
(\partial _{i}\vec{n})^{2}+(\partial _{i}\vec{m})^{2} &=&2(\vec{n}\partial
_{i}\vec{m})^{2}+(\partial _{i}\vec{l})^{2},  \label{rel2} \\
\varepsilon _{pqs}l_{p}(\partial _{i}l_{q})(\partial _{j}l_{s}) &=&(\partial
_{i}n_{p})(\partial _{j}m_{p})-(\partial _{i}m_{p})(\partial _{j}n_{p}).
\label{rel3}
\end{eqnarray}

To obtain the free energy density of the London approximation we substitute $%
\vec{\psi}$ \ in the form eq.(\ref{phaseI}) into the gradient part, eq.(\ref
{Fgrad}), of the total free energy functional and make use of eq.(\ref{rel1}%
) and eq.(\ref{rel2}). After some algebra we get:

\begin{equation}
F_{L}=\frac{1}{2}\left( \partial _{i}\vec{l}\right) ^{2}+\left( \vec{n}%
\partial _{i}\vec{m}-a_{i}\right) ^{2}+b_{i}^{2}.  \label{main functional}
\end{equation}
Varying energy functional with respect to vector potential $\vec{a}$ one
obtains the supercurrent equation: 
\begin{equation}
n_{p}\vec{\nabla}m_{p}-\vec{a}=\vec{\nabla}\times \left( \vec{\nabla}\times 
\vec{a}\right) =\vec{j},  \label{J-eq}
\end{equation}
where the Maxwell equation was used. Eq.(\ref{J-eq}) shows that the
superconducting velocity (in units of $\hbar /m^{\ast }$) is given by 
\begin{equation}
n_{p}\vec{\nabla}m_{p}=-\vec{\nabla}\vartheta .  \label{SCphase}
\end{equation}
Thus, the angle $\vartheta ,$ which specifies the position the pair of
perpendicular unit vectors $\vec{n}$ and $\vec{m}$ in the plane normal to
vector $\vec{l},$ takes the role of superconducting phase in the present
case (see Fig.1). Other field equations are most easily obtained by
considering $F_{L}(\vec{l},\vec{n},\vec{m})$ as a functional of $\vec{l}$
and $\vec{n}$ only and performing conditional variation with constraints $%
\vec{l}\cdot $ $\vec{n}=0,$ $\vec{l}^{2}=\vec{n}^{2}=1.$ This procedure
yields independent equation for $\vec{l}$: 
\begin{equation}
\Delta \vec{l}-\vec{l}(\vec{l}\cdot \Delta \vec{l})+2j_{k}(\vec{l}\times
\partial _{k}\vec{l})=0.  \label{OP-eq}
\end{equation}

\section{Topological classification of solutions in London Approximation}

In this subsection we develop a classification scheme for the finite energy
solutions to our model in the London approximation derived above. The main
results is that the London equation, eq.(\ref{J-eq}) and eq.(\ref{OP-eq}),
in the presence of the magnetic flux admit nonsingular topologically stable
solutions. This class of solutions contains cylindrically symmetric ones.

\subsection{\protect\bigskip General topological analysis}

Let us consider boundary conditions for a superconductor which extends over
the whole space. The free energy density eq.(\ref{main functional}) is
positive definite and contains $\vec{B}^{2}$ term. It follows that magnetic
field vanishes at spatial infinity. Then one has to specify the triad $\vec{n%
},$ $\vec{m},$ $\vec{l}$ at different distant points. The corresponding
(first) homotopy group of vacuum manifold is $\pi _{1}(SO(3))=Z_{2}$ \cite
{Salomaa}. It yields a classification of finite energy solutions into two
topologically distinct classes. This classification is too weak, however,
because it doesn't guarantee nontrivial flux penetrating the plane. We will
see that configurations having both ''parities'' are of interest.

In the presence of the magnetic flux, the configurations are further
constrained due to the flux quantization condition. The vacuum manifold is
naturally divided into $SO(3)\rightarrow SO(2)\otimes S_{2},${\sl \ }where
the $S_{2}$ is the direction of $\vec{l}$ and the $SO(2)$ is the
superconducting phase $\vartheta $ defined in eq.(\ref{SCphase}). For given
number of flux quanta $N\equiv \Phi /\Phi _{0}$, the phase $\vartheta $
makes $N$ winds at infinity, see Fig.2. The first homotopy group of this
part is therefore fixed: $\pi _{1}(SO(2))=Z.$ If, in addition, vector $\vec{l%
}$ is fixed throughout the volume of a superconductor there is no way to
avoid singularity in the phase $\vartheta $. It becomes ill defined at some
point and, accordingly, the modulus of the order parameter have to vanish
there. Destruction of the superconducting state takes place in rather small
area, especially for large $\kappa .$ Thus, we arrive at usual picture of
Abrikosov vortex.

However, the general requirement that a solution has finite energy is much
weaker. It tells us that the direction of $\vec{l}$ should be fixed only at
infinity This follows from the presence of $\left( \partial _{i}\vec{l}%
\right) ^{2}$ term in $F_{L}$ (see eq.(\ref{main functional})) which cannot
be ''gauged away'' as the corresponding term for the $SO(2)$ part. A
relevant homotopy group is $\pi _{2}(S_{2})=Z.$ The second homotopy group
appears because constancy of $\vec{l}$ at infinity (say, up) effectively
''compactifies'' the two dimensional physical space into $S_{2}.$ One can
have topologically nontrivial configuration, skyrmions, which are markedly
different from vortices. Unit vector $\vec{l}$ can nontrivially wind towards
the center of the texture. New topological number $Q$ should be introduced 
\cite{Rajaraman}: 
\begin{equation}
Q=\frac{1}{8\pi }\int \varepsilon _{ij}\,\vec{l}\,\left( \partial _{i}\vec{l}%
\times \partial _{j}\vec{l}\right) dS.  \label{sk-number}
\end{equation}
Configurations of the order parameter field with topological number $Q=1$\
have vector $\vec{l}$\ flipping its direction from up to down (or from down
to up) until it reaches the center of the texture from an infinitely remote
point (see Fig.2).

To summarize, configurations fall into classes characterized by two integers 
$N$ and $Q.$ The ''parity'' of the more general topological analysis is just 
$Q=N(%
\mathop{\rm mod}%
2).$ Due to the presence of two topological numbers an interesting
possibility arises. There exists topologically nontrivial configuration that
preserves the modulus of the order parameter (see eq.(\ref{phaseI})) at
every point. We call these regular solutions magnetic skyrmions. For them
these two topological numbers are related to each other. We find this
relation integrating the supercurrent equation, eq.(\ref{J-eq}) along a
remote contour and using of the identity eq.(\ref{rel3}): 
\begin{equation}
Q=N/2.  \label{condition}
\end{equation}

The lowest energy solution within the London approximation corresponds to $%
N/2=Q=\pm1$.

\subsection{\protect\bigskip Cylindrically symmetric magnetic skyrmions.}

In the class of solution $N/2=Q=-1$ there are ones possessing cylindrical
symmetry. We will describe them in polar coordinates, $\rho $ and $\varphi $%
. The triad $\vec{n},$ $\vec{m},$ $\vec{l}$ has form: 
\begin{eqnarray}
\vec{l} &=&\vec{e}_{z}\cos \Theta (\rho )+\vec{e}_{\rho }\sin \Theta (\rho ),
\nonumber \\
\vec{n} &=&\left( \vec{e}_{z}\sin \Theta (\rho )-\vec{e}_{\rho }\cos \Theta
(\rho )\right) \sin \varphi +\vec{e}_{\varphi }\cos \varphi
\label{cyl-sym-skyrmion} \\
\vec{m} &=&\left( \vec{e}_{z}\sin \Theta (\rho )-\vec{e}_{\rho }\cos \Theta
(\rho )\right) \,\cos \varphi -\vec{e}_{\varphi }\sin \varphi ,  \nonumber
\end{eqnarray}
where $\Theta $ is the azimuthal angle of $\vec{l}$ (see Fig.1). This choice
corresponds to the situation when the pair of perpendicular vectors $\vec{n}$
and $\vec{m}$ winds twice as a distant circle on Fig.2 is completed. Due to
cylindrical symmetry of the solution in question function $\Theta (\rho )$
satisfies boundary conditions $\Theta =\pi $ at $\rho =0$ and $\Theta =0$ at 
$\rho \rightarrow \infty $.

The free energy of the magnetic skyrmion per unit length takes form: 
\begin{eqnarray}
\epsilon _{ms} &=&\varepsilon _{s}+\varepsilon _{cur}+\varepsilon _{mag}
\label{L-energy-rot-sym} \\
\varepsilon _{s} &\equiv &\int \rho d\rho \left[ \frac{1}{2}\left( \frac{%
d\Theta }{d\rho }\right) ^{2}+\frac{\sin ^{2}\Theta }{2\rho ^{2}}\right] \\
\varepsilon _{cur} &\equiv &\int \rho d\rho \left( \frac{1+\cos \Theta }{%
\rho }+a\right) ^{2} \\
\varepsilon _{mag} &\equiv &\int \rho d\rho B^{2}=\int \rho d\rho \left( 
\frac{a}{\rho }+\frac{da}{d\rho }\right) ^{2},
\end{eqnarray}
where energy is measured in units of $\epsilon _{0}=\left( \frac{\Phi _{0}}{%
4\pi \lambda }\right) ^{2}.$ The first part $\epsilon _{ms}$ is the same as
in standard nonlinear $\sigma -$ model without magnetic field \cite
{Rajaraman}. The second term $\varepsilon _{cur}$ is analogous to the
supercurrent contribution in the London approximation of the usual
superconductor \cite{Tinkham}. The third term is the magnetic energy. Eq. (%
\ref{L-energy-rot-sym}) shows that a singularity at $\rho =0$ is absent
(integrand converges) since $1+\cos \Theta (0)=0.$

Actual distribution of magnetic field and order parameter in this case can
be found from the following system of equations: 
\begin{eqnarray}
\Theta ^{\prime \prime }+\frac{1}{\rho }\Theta ^{\prime } &=&-\frac{\sin
\Theta }{\rho }\left( \frac{2+\cos \Theta }{\rho }+2a\right) ,
\label{rot-eq1} \\
a^{\prime \prime }+\frac{a^{\prime }}{\rho }-\frac{a}{\rho ^{2}}-a &=&\frac{1%
}{\rho }\left( 1+\cos \Theta \right) .  \label{rot-eq2}
\end{eqnarray}
In the next section we solve this equation.

\section{Magnetic skyrmion solution}

\subsection{Blow up of single skyrmion by magnetic field}

The general form of the solution\ of eq.(\ref{rot-eq1}) and eq.(\ref{rot-eq2}%
) is given on Fig. 2. The orientation of the unit vector $\vec{l}$ (solid
arrows) forms a skyrmion of $SO(3)$ invariant $\sigma $- model \cite
{Rajaraman}. The phase $\vartheta $ makes two rounds at infinity (clock
inside small circles on the ''infinitely remote'' circle). If magnetic field
were absent there are infinitely many degenerate solutions

\begin{equation}
\Theta _{s}(\rho )=2\arctan (\delta /\rho )  \label{skyrmsol}
\end{equation}
which have the same energy $\varepsilon =2$ for any size of the skyrmion $%
\delta $. The skyrmion of nonlinear $\sigma $- model possesses a scale
invariance. This degeneracy in various physical problem is lifted by
perturbations. In some physical situations the skyrmion is stabilized by
four derivative terms \cite{Rajaraman}, sometimes it shrinks and sometimes
blows up. In the present context magnetic field lifts the degeneracy and we
prove below that the skyrmion blows up. Of course is there are many
skyrmions present their repulsion with stabilize the system. This is
discussed in the next subsection.

To prove that the skyrmion blows up, we explicitly construct variational
configurations and show that as size of these configurations increases, the
energy is reduced to a value arbitrarily close to the absolute minimum of $%
\varepsilon _{ms}=2$.

The first term in the energy eq.(\ref{L-energy-rot-sym}) $\varepsilon _{s}$
is the usual expression for the energy of the skyrmion. It is bound from
below by the energy of usual skyrmion $\varepsilon =2$. To construct a
variational configuration for $\Theta $, we pick up one of these solutions
eq.(\ref{skyrmsol}) of certain size $\delta $. The second term \ $%
\varepsilon _{cur}$, the ''supercurrent'' contribution is positive definite.
Therefore its minimum cannot be lower then zero. One still can maintain the
zero value of this term when the field $\Theta $ is a skyrmion. Assuming
this one gets the relation between $a$ and $\Theta :a(\rho )=-\frac{1+\cos
\Theta }{\rho }=-\frac{2\rho }{\rho ^{2}+\delta ^{2}}.$ The magnetic field
contribution (which is also positive definite) for such a vector potential
is: $\varepsilon _{mag}=\frac{8}{3\delta ^{2}}$. To sum up, the energy of
the configuration is $\varepsilon =2+\frac{8}{3\delta ^{2}}$. It is clear
that when $\delta \rightarrow \infty $, we obtain energy arbitrarily close
to the lower bound of $\varepsilon =2.$ The skyrmion therefore blows up.

We also solved eqs.(\ref{rot-eq1})--(\ref{rot-eq2}) numerically on the
segment of $\rho $ from 0 to a cutoff $\rho _{\max }$ with boundary
conditions $b|_{\rho =\rho _{\max }}=0$ and $\left( \Theta ^{\prime }+\Theta
/\rho \right) |_{\rho =\rho _{\max }}=0$. The second boundary condition
allows us to approach the correct asymptotic behavior of $\Theta $ at
infinity $\sim 1/\rho $ \ which follows from eqs.(\ref{rot-eq1})--(\ref
{rot-eq2}). The results for the distribution of magnetic field for $\rho
_{\max }$ ranging from 50 to 600 are presented in Fig. 3. One clearly sees
that as the cutoff increases the magnetic field at the center $\rho =0$
decreases and the flux spreads out over larger area. This is in accord with
the variational proof above.

\subsection{\protect\bigskip Skyrmion lattice and $H_{c1}$}

Skyrmions repel each other, as we will see shortly, and therefore form a
lattice. Since they are axially symmetric objects, the interaction is
axially symmetric and hexagonal lattice is expected (see Fig.4). Assume that
lattice spacing is $a_{\bigtriangleup }$. At the boundaries of the hexagonal
unit cells the angle $\Theta $ is zero, while at the centers it is $\pi $.
Magnetic field $b$ is continuous on the boundaries. Therefore, to analyze
magnetic skyrmion lattice we should solve the equations (\ref{rot-eq1})-(\ref
{rot-eq2}) on the unit cell with such a boundary conditions demanding that
two units of flux pass through the cell (by adjusting the value of magnetic
field on the boundary). We approximate the hexagonal unit cell by a circle
of radius $R=\frac{3^{1/4}}{\sqrt{2\pi }}a_{\bigtriangleup }$ having the
same area, Fig.4.

We performed such calculations for $R$ from $R=5$ till $R=600$ using finite
elements method. The result is presented in Fig.5. The energy per unit cell
is described well in a wide range of $R$ (deviation at $R=10$ is $1\%$) by
an approximate expression 
\begin{equation}
\varepsilon _{cell}=2+\frac{5.62}{R}  \label{unit-cell-energy}
\end{equation}
The dominant constant contribution to the energy at large $\ R$ comes like
in the analytical variational state above from the first term $\varepsilon
_{s}$ in the integrand of eq.(\ref{L-energy-rot-sym}) . The contribution to
the energy eq.(\ref{L-energy-rot-sym}) from the supercurrent term $%
\varepsilon _{cur}$ is small for large $R$ but becomes significant at denser
lattices. The third term, magnetic energy $\varepsilon _{mag}$ yields a
small deviation of magnetic skyrmion energy from $2$ at large $R$.

Profile of the angle $\Theta (\rho )$ and of the magnetic field $b$ are
depicted on Fig.6a and 6b respectively. Radius of the circular cell $R$
varies from $20\lambda $ to $300\lambda $. On Fig.6a smaller value of $R$
corresponds to a lower curve. Small $\rho $ asymptotics of the solution up
to $\rho ^{3}$ terms read: 
\begin{eqnarray*}
\Theta (\rho ) &\rightarrow &\pi +c\rho \left[ 1+\frac{\rho ^{2}}{8}\left(
b(0)+\frac{c^{2}}{3}\right) \right] , \\
a(\rho ) &\rightarrow &\frac{b(0)}{2}\rho +\frac{\rho ^{3}}{16}\left(
b(0)+c^{2}\right) ,
\end{eqnarray*}
where $c$ and $b(0)$ are constants to be determined by numerical
integration. Most of the flux goes through the region where the vector $\vec{%
l}$ is oriented upwards. In other words, the magnetic field is concentrated
close to the center of a magnetic skyrmion.

The value of $h_{c1}(R\rightarrow \infty )=\varepsilon _{ms}(R\rightarrow
\infty )/4$ for a triplet superconductor filling the whole space is equal to 
$1/2.$ In physical units this result reads: 
\begin{equation}
H_{c1}=\frac{\Phi _{0}}{4\pi \lambda ^{2}}.  \label{Hc1skyrm}
\end{equation}
It is quite different from $H_{c1}$ of conventional (s-wave) superconductors
where an additional factor $\log \kappa $ is present. Line energy of
Abrikosov vortices $\varepsilon _{v}$ for the present model was calculated
numerically (beyond London approximation) in \cite{Knigavko2}. For $\kappa
=20$ and \ $50$ we obtain $2\varepsilon _{v}/\varepsilon _{ms}\approx 3.5$
and $4.4$ respectively. Therefore we expect that the lower critical field of 
$UPt_{3}$ is determined by magnetic skyrmions.

\subsection{Magnetization of the skyrmion lattice}

If $h>h_{c1}$ the external magnetic field enforces a definite value of
magnetic flux through a sample. Magnetic skyrmions, being topological
objects, carry quantized magnetic flux and their number in the sample is
determined by the average magnetic induction $b,$ similarly to the case of
vortices. Energy of magnetic skyrmions as function of $R$ eq.(\ref
{unit-cell-energy}) actually determines the interaction between them.
However, magnetic skyrmions, contrary to vortices, are extended objects and
their linear size $R$ is also determined by the number of them in the sample.

To qualitatively estimate the magnetization curve produced by ''skyrmion
mixed state'' we make use of the unit cell energy obtained in the previous
section. The Gibbs energy density of the sample of volume $V=S\times L,$
where $S$ is the transverse area and $L$ is longitudinal extension, is given
in dimensionless units eq.(\ref{units}) by

\begin{equation}
G(b)=\frac{N\varepsilon _{cell}L}{V}-\,2bh=\frac{b}{2}\left( 2+5.62\frac{%
\sqrt{b}}{2}\right) -2bh.  \label{Gibbs energy}
\end{equation}
The second equality follows the facts that magnetic induction $b$ is related
both to the number of magnetic skyrmions $N=Sb/2$ and to the size of the
magnetic skyrmion defined above $R^{2}=4/b.$ Minimization of eq.(\ref{Gibbs
energy}) with respect of $b$ yields: 
\begin{equation}
b\simeq 0.225\left( \frac{h}{h_{c1}}-1\right) ^{2},\;\;\;h\geq h_{c1}=\frac{1%
}{2}.  \label{magnetization}
\end{equation}

Eq.(\ref{magnetization}) shows that a skyrmion lattice is characterized by
zero slope of magnetization curve at $h_{c1}$, in contract to the infinite
slope for the magnetization curve associated with a vortex lattice. This
circumstance provides a tool in the experimental search for the triplet
superconductivity with approximate SO(3) symmetry. Our results agree well
with the earlier work of Burlachkov {\it et al }\cite{Burlachkov} who also
obtained zero slop of the magnetization at $h_{c1}$ for a stripe $\vec{l}$
texture which might arise in the case of very high anisotropy of effective
mass tensor $m^{\ast }$ (see eq.(\ref{Free energy})).

\section{Influence of $SO(3)$ breaking terms}

In this section we consider influence of an $SO(3)$ symmetry breaking terms
on skyrmion lattice. List of these terms was given in section IIB eqs.(\ref
{deva}--\ref{deve}). The perturbations are not expected to affect the
existence of topological solitons - just modify their energy. When the
coefficient of a breaking term becomes of order 1, the soliton might
disappear, although it is not necessary. We study in detail the influence of
Zeeman term eq.(\ref{devd}). The choice is motivated by our previous study
of possible spontaneous vortex state in a new bulk perovskite superconductor 
$Sr_{2}YRu_{1-x}Cu_{x}O_{6}$ \cite{Knigavko2}.

This compound has very unusual magnetic properties and is suspected to be a $%
p$-wave superconductor for the following reasons\cite{Wu}. At the
temperature of about 60K, at which superconductivity sets in, these
materials begin to exhibit basic ferromagnetic properties like hysteresis
loop. Experimental observation of a positive remanence suggests existence of
spontaneous magnetization in the absence of an external magnetic field.
Exact overlap of superconductivity and ferromagnetism lead us to consider an
isotropic triplet model eqs.(\ref{Free energy})--(\ref{Fgrad}) in nonunitary
phase with spontaneous time reversal symmetry breaking. In this case, a
direct spin coupling of the condensate to a magnetic field 
\begin{equation}
\mu \vec{S}\cdot \vec{B}=\frac{e^{\ast }\hbar }{2m^{\ast }c}g\vec{S}\cdot 
\vec{B}  \label{break-term}
\end{equation}
becomes relevant. In what follows this coupling will be referred to as
Zeeman-like coupling and characterized by dimensionless parameter $g$. For
sufficiently large values of $g$ energetics of the triplet superconductor
changes considerably. There exists a critical value $g_{c1}=1$ above which
the mixed state might respond on an external magnetic filed
ferromagnetically and, on the other hand, in the presence of an external
magnetic the field mixed state might occur even for temperatures above $%
T_{c} $ \cite{Knigavko2}. For larger Zeeman-like coupling, $g>g_{c2}\approx
\log \kappa $, vortex energy becomes negative. Spontaneous vortex phase
appears at $H=0$ and exists for arbitrarily large magnetic field. Meissner
phase, therefore, completely disappears. Vortices become thinner when $H$
grows. The structure of the vortex core is markedly different from the usual
one.

Our analysis in \cite{Knigavko2} was entirely based on the simplest possible
topological objects: vortices of the usual type. Value of $\kappa $ for the
materials of Wu {\it et al }\cite{Wu} are estimated to be quite large and,
consequently, vortices should be heavy compared to magnetic skyrmions.
Spontaneously magnetized skyrmion lattice can also occur, as in the previous
case of vortices of usual type. Values of $g$ required to obtain spontaneous
vortex state $g=\log \kappa $ were very high and made the scenario
questionable. This value is lowered to $g\sim 1$ for magnetic skyrmion
lattice.

The free energy per unit length for a single magnetic skyrmion now has form:

\begin{eqnarray}
{\cal F}_{L} &=&\int \rho d\rho \left[ \frac{1}{2}\left( \frac{d\Theta }{%
d\rho }\right) ^{2}+\frac{\sin ^{2}\Theta }{2\rho ^{2}}+\left( \frac{1+\cos
\Theta }{\rho }+a\right) ^{2}\right. \\
&&\left. +\left( \frac{a}{\rho }+\frac{da}{d\rho }\right) ^{2}-g\left( \frac{%
a}{\rho }+\frac{da}{d\rho }\right) \cos \Theta \right] .  \nonumber
\end{eqnarray}

The equations read:

\begin{eqnarray}
\Theta ^{\prime \prime }+\frac{1}{\rho }\Theta ^{\prime } &=&-\frac{\sin
\Theta }{\rho }\left( \frac{2+\cos \Theta }{\rho }+2a\right) +g\sin \Theta
\left( a^{\prime }+\frac{a}{\rho }\right) , \\
a^{\prime \prime }+\frac{a^{\prime }}{\rho }-\frac{a}{\rho ^{2}}-a &=&\frac{1%
}{\rho }\left( 1+\cos \Theta \right) -\frac{g}{2}\Theta ^{\prime }\sin
\Theta .
\end{eqnarray}

We use the same boundary conditions as that for the case of isolated
magnetic skyrmion at $g=0$ (see Sec.IV A) Calculations were performed both
for positive and negative values of $g.$ Plot of the energy of the magnetic
skyrmion as a function of $g$ is presented in Fig. 7. The characteristic
feature of this dependence are a maximum near $g=0$. Profiles of the
magnetic field $b(\rho )$ for different $g$ of both signs are presented in
Fig 8. Zeeman interaction strongly influences behavior of $b(\rho )$ near
the center of the magnetic skyrmion and in quite different manner for
positive and negative $g$. Note, however, that as $|g|$ increases behavior
of the function changes significantly in the interval of $\rho $ from the
origin up to only some limiting value, after which it remains approximately
the same for different $g$. Thus we observe that nonzero $g$ actually
introduces new length scale in the problem. Changes in the profile of $%
\Theta (\rho )$ with $g$ are less pronounced and are not displayed.

\section{Discussion}

In this paper we performed topological classification of solutions in $SO(3)$
symmetric Ginzburg - Landau free energy. This model with addition of very
small symmetry breaking terms describes heavy fermion superconductor $%
UPt_{3} $ and possibly other triplet superconductors. A new class of
topological solutions in weak magnetic field carrying two units of magnetic
flux was identified. These solutions, magnetic skyrmions, are nonsingular
(do not have singular core like Abrikosov vortices). They repel each other
as $1/r$ at distances much larger then magnetic penetration depth $\lambda $
forming relatively robust triangular lattice. At lattice spacings much
larger then $\lambda $ their energy is reduced by a factor of the order of $%
\log \kappa $ as compared to the usual Abrikosov vortex solutions and
therefore dominate the magnetic properties for strongly type II
superconductors. The lower critical magnetic field $H_{c1}=$ $\frac{\Phi _{0}%
}{4\pi \lambda ^{2}}$ is reduced correspondingly by a factor $2\log \kappa $.

Magnetization near $H_{c1}$ instead of sharply rising with infinite
derivative increases gradually as $(H-H_{c1})^{2}$. This agrees very well
with the experimental results for $UPt_{3}$, see Fig.1 of \cite{Knigavko1}.
For fields higher then several $H_{c1}$ London approximation is not valid
anymore since magnetic skyrmions will start to overlap. At distances between
fluxons of order $\lambda $ (or at the field $H_{c1}^{\prime }\sim
H_{c1}\cdot 2\log \kappa $) one expects that ordinary Abrikosov vortices,
which carry one unit of magnetic flux, become energetically favorable. The
usual vortex picture has indeed been observed at high fields by Yaron{\it \
et al}\cite{Yaron}. Curiously, our result on magnetization is similar to
conclusions of Burlachkov {\it et al.}\cite{Burlachkov} who investigated
stripe-like (quasi one dimensional) spin textures in triplet
superconductors. Magnetic skyrmions are quite stable objects and they are
not destroyed by small perturbations of exact $SO(3)$ symmetry of the
original model eqs.(\ref{Free energy}--\ref{Fgrad}). Moreover, deformed
magnetic skyrmions might exist even at large deviations from exact $SO(3)$
symmetry. We demonstrated this including Zeeman like interaction eq.(\ref
{break-term}).

Let us list below the experimental features which can allow identification
of the magnetic skyrmions lattice.

1. The lower critical filed is significantly smaller then usually expected.
For such strongly type II superconductors as $UPt_{3}$, $Sr_{2}RuO_{4}$ or $%
Sr_{2}YRu_{1-x}Cu_{x}O_{6}$ $\ $with $\kappa \sim 50\div 70$ the reduction
amounts 8 times. Although $H_{c1}$ is expected to be very small (less then $%
1Gauss$) it is still measurable.

2. Magnetization above $H_{c1}$, but below crossover to Abrikosov vortex
lattice $H_{c1}^{\prime }\sim \frac{\Phi _{0}}{2\pi \lambda ^{2}}\log \kappa 
$ is markedly distinct from the usual one due to long range nature of the
magnetic skyrmions.

3. Unit of flux quantization is different: $2\Phi _{0}$.

4. Magnetic field profile is different: no exponential drop even at very
sparse lattices.

5. Superfluid density $|\overrightarrow{\psi }|^{2}$ is almost constant
throughout the mixed state. There is no normal cores of the fluxons. This
can be tested using scanning tunneling microscopy technique.

6. Due to the fact that there is no small normal core where usually
dissipation and pinning take place, one expects that pinning effects are
greatly reduced. Correspondingly critical current should be very small.

7. The vortex lattice in the region around $H_{c1}$ can melt into so called
lower field vortex liquid due to thermal fluctuations \cite{Nelson}. The
melting of usual Abrikosov vortex lattice is easy even in not very strongly
fluctuating superconductors because interaction between Abrikosov vortices
is exponentially small. It is not so for magnetic skyrmions. Due to their
long range $\ 1/r$ interaction the lattice is more robust and therefore no
melting is expected.

\section{Acknowledgments}

The authors are grateful to B. Maple for discussion of results of Ref. 13,
to L. Bulaevskii, T.K. Lee, H.C. Ren, J. Sauls and M.K. Wu for discussions
and to A. Balatsky for hospitality in Los Alamos. The work is supported by
NSC, Republic of China, through contract \#NSC86-2112-M009-034T.

\bigskip

\eject
\centerline{\large \bf List of Figures}

\begin{enumerate}
\item[Fig. 1]  Definition of angles $\vartheta $ and $\Theta .$ Unit vectors 
$\vec{l},$ $\vec{n},$ $\vec{m}$ constitute a triad of perpendicular vectors
in the spin space. $\vartheta $ is the superconducting phase defined in eq.(%
\ref{SCphase}).

\item[Fig. 2]  Configuration of a magnetic skyrmion with $Q=-1$. Solid
arrows represent $\vec{l}$ field while ''clocks'' show that phase $\vartheta 
$ rotates twice clockwise as a round on the remote contour is completed.

\item[Fig. 3]  Magnetic field of the isolated magnetic skyrmion. Distance
from the center $\rho $ varies from 0 to a cutoff $\rho _{\max }$ with
boundary conditions $\left( \Theta ^{\prime }+\Theta /\rho \right) |_{\rho
=\rho _{\max }}=0,b|_{\rho =\rho _{\max }}=0$ imitating the infinite domain. 
$\rho _{\max }=40\lambda $ for the lowest curve and $600$ for the uppermost
one.

\item[Fig. 4]  A fragment of the magnetic skyrmion lattice. For numerical
calculations we approximate symmetric unit cell by the disk of the same
area: $R=\frac{3^{1/4}}{\sqrt{2\pi }}a_{\bigtriangleup }$.

\item[Fig. 5]  Energy of the unit cell of the magnetic skyrmion lattice.
Dots are numerical values for different $R.$ Line is the fit of eq.(\ref
{unit-cell-energy}).

\item[Fig. 6]  Numerical solution of GL equations in London approximation
for a unit cell of the magnetic skyrmion lattice. Radius of the circular
cell $R$ varies from $20\lambda $ to $300\lambda $. a) Angle $\Theta $ as a
function of the distance $\rho $ from the center of the cell. b) Magnetic
field $b$ as a function of the distance $\rho $ from the center of the cell.
A smaller $R$ corresponds to a lower curve.

\item[Fig. 7]  Energy of the isolated magnetic skyrmion as a function of
dimensionless Zeeman coupling $g$ for $R/\lambda =300.$

\item[Fig. 8]  Magnetic field of the isolated magnetic skyrmion as a
function of distance from the center $\rho $ for different Zeeman coupling
and for the case of $\rho _{\max }=300\lambda $ (see Fig. 3 caption). a) $%
g=0,.5,.7,.9,1.,1.1$. b) $g=0,-.5,-.7,-.9,-1.,-1.1.$ In both cases a smaller 
$|g|$ corresponds to a lower at $\rho =0$ curve.
\end{enumerate}

\eject
\begin{figure}[htp]
\epsfig{figure=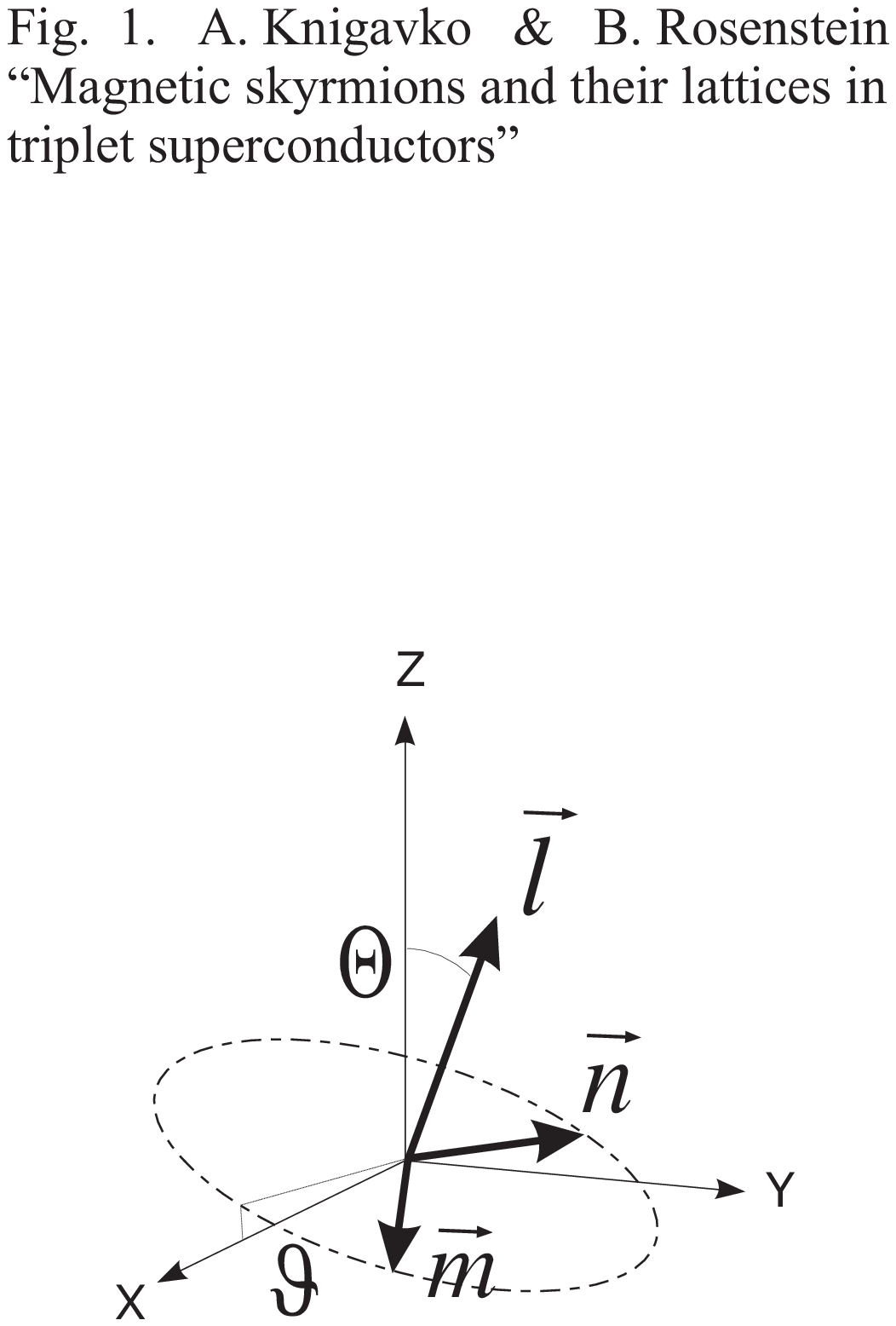,height=7in,width=5.0in,angle=-0}
\end{figure}

\eject
\begin{figure}[htp]
\epsfig{figure=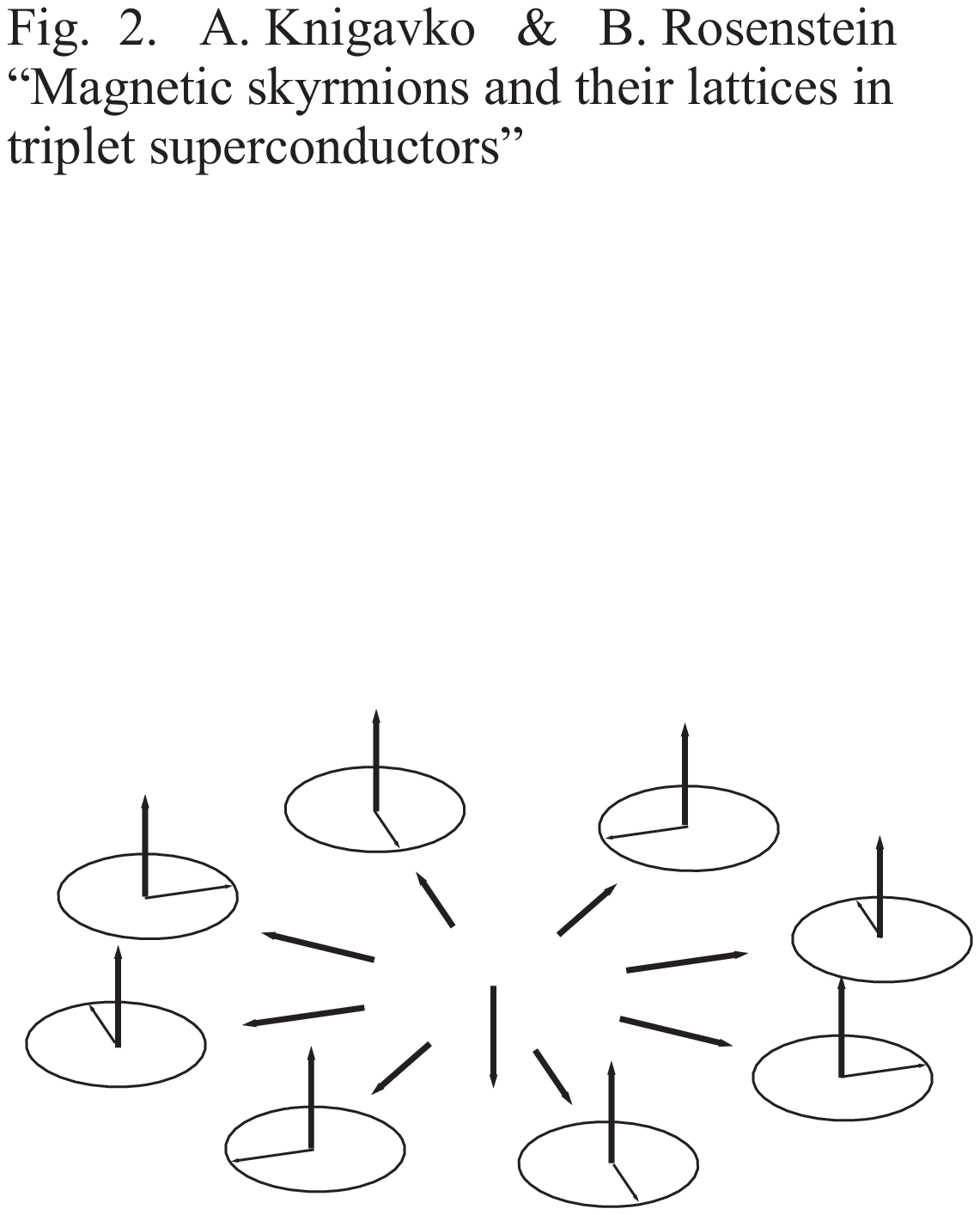,height=7in,width=5.0in,angle=-0}
\end{figure}

\eject
\begin{figure}[htp]
\epsfig{figure=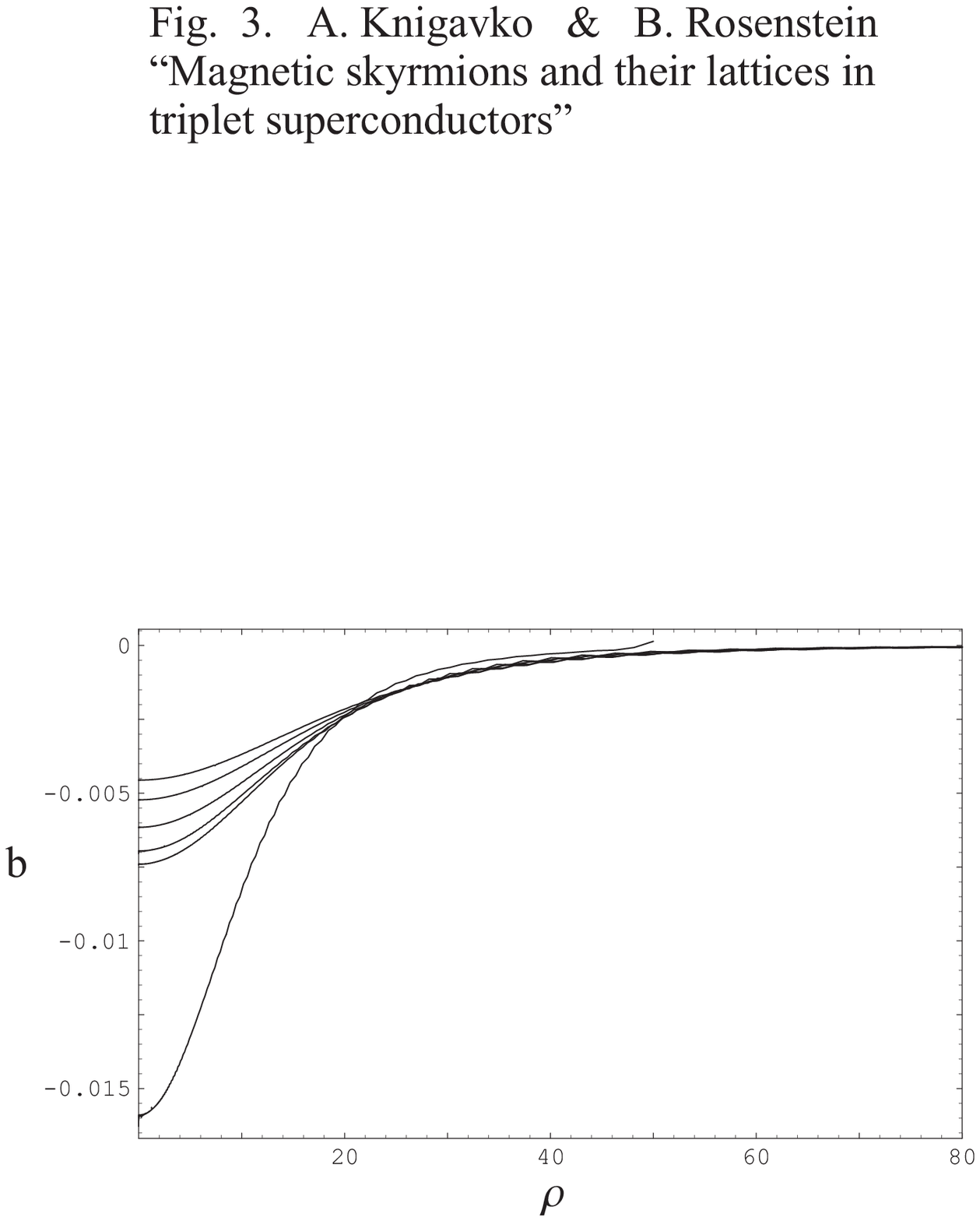,height=7in,width=5.0in,angle=-0}
\end{figure}

\eject
\begin{figure}[htp]
\epsfig{figure=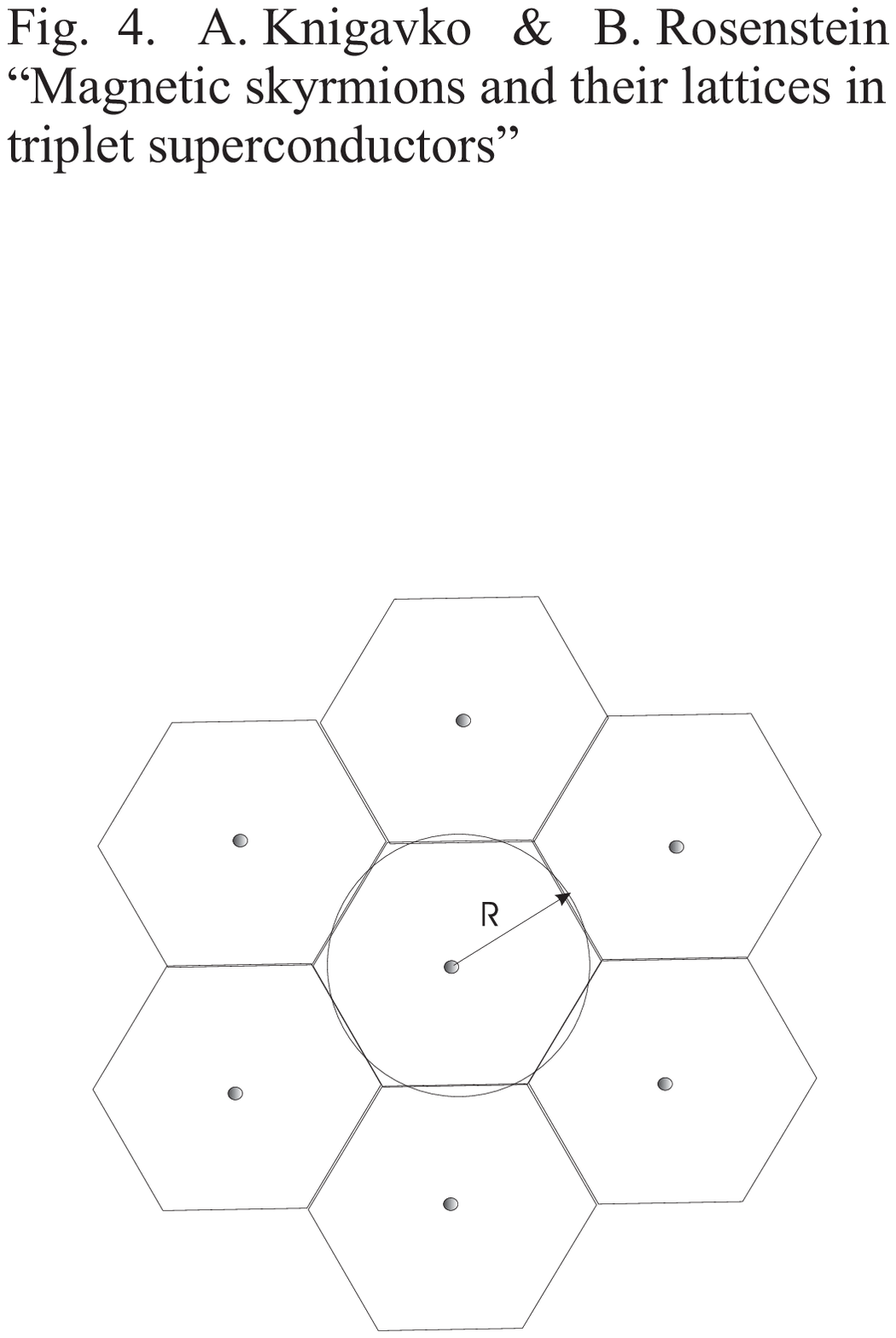,height=7in,width=5.0in,angle=-0}
\end{figure}

\eject
\begin{figure}[htp]
\epsfig{figure=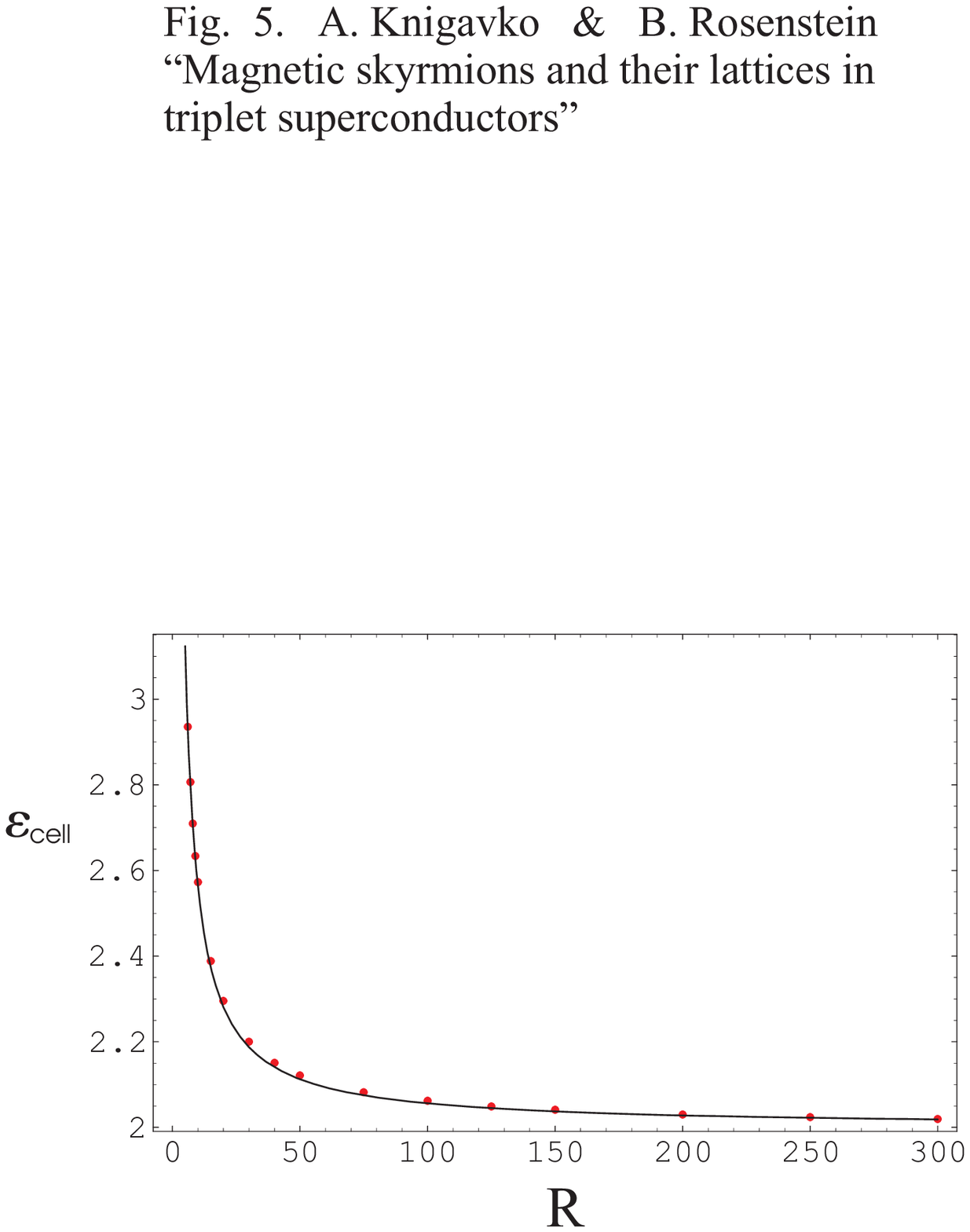,height=7in,width=5.0in,angle=-0}
\end{figure}

\eject
\begin{figure}[htp]
\epsfig{figure=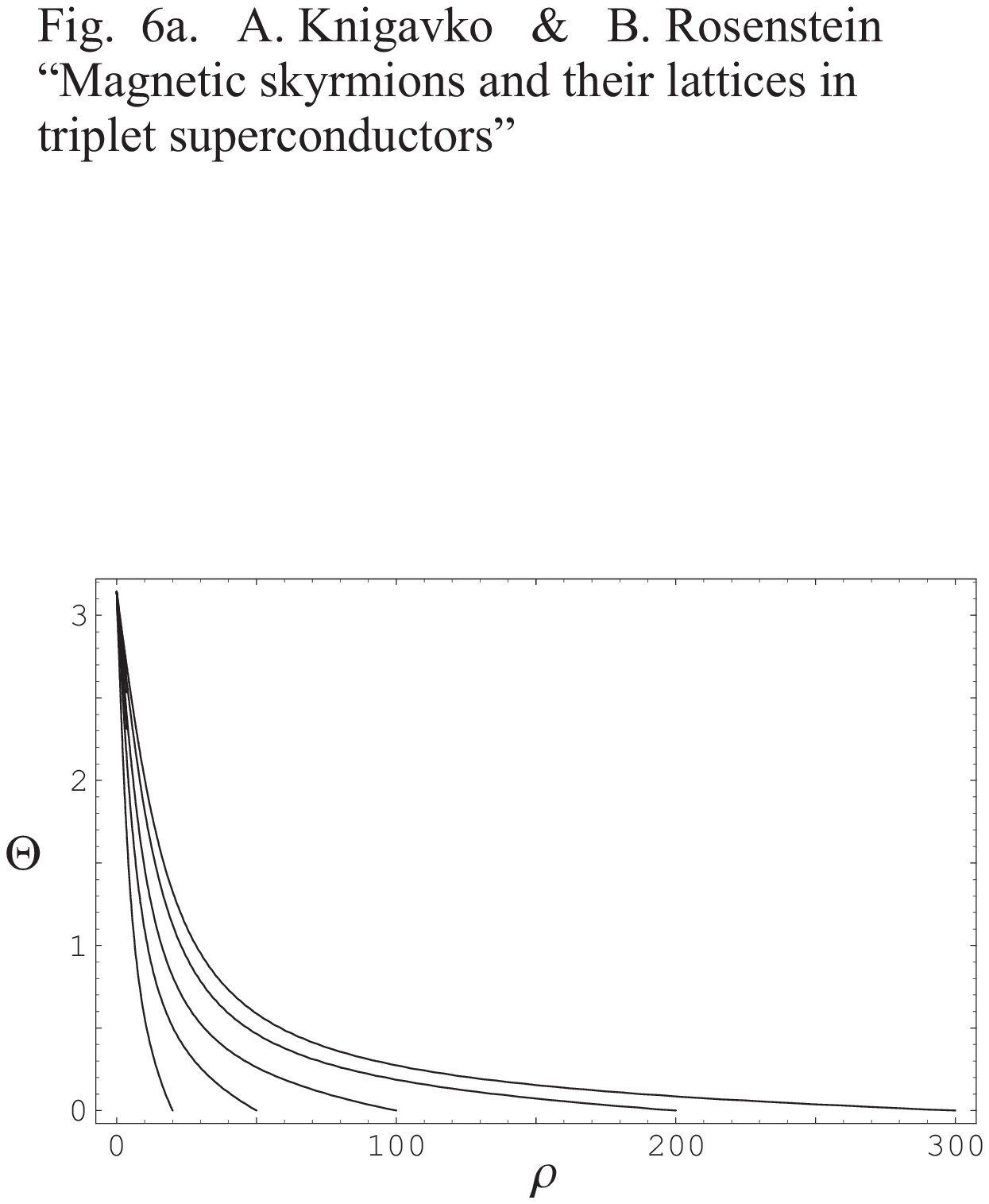,height=7in,width=5.0in,angle=-0}
\end{figure}

\eject
\begin{figure}[htp]
\epsfig{figure=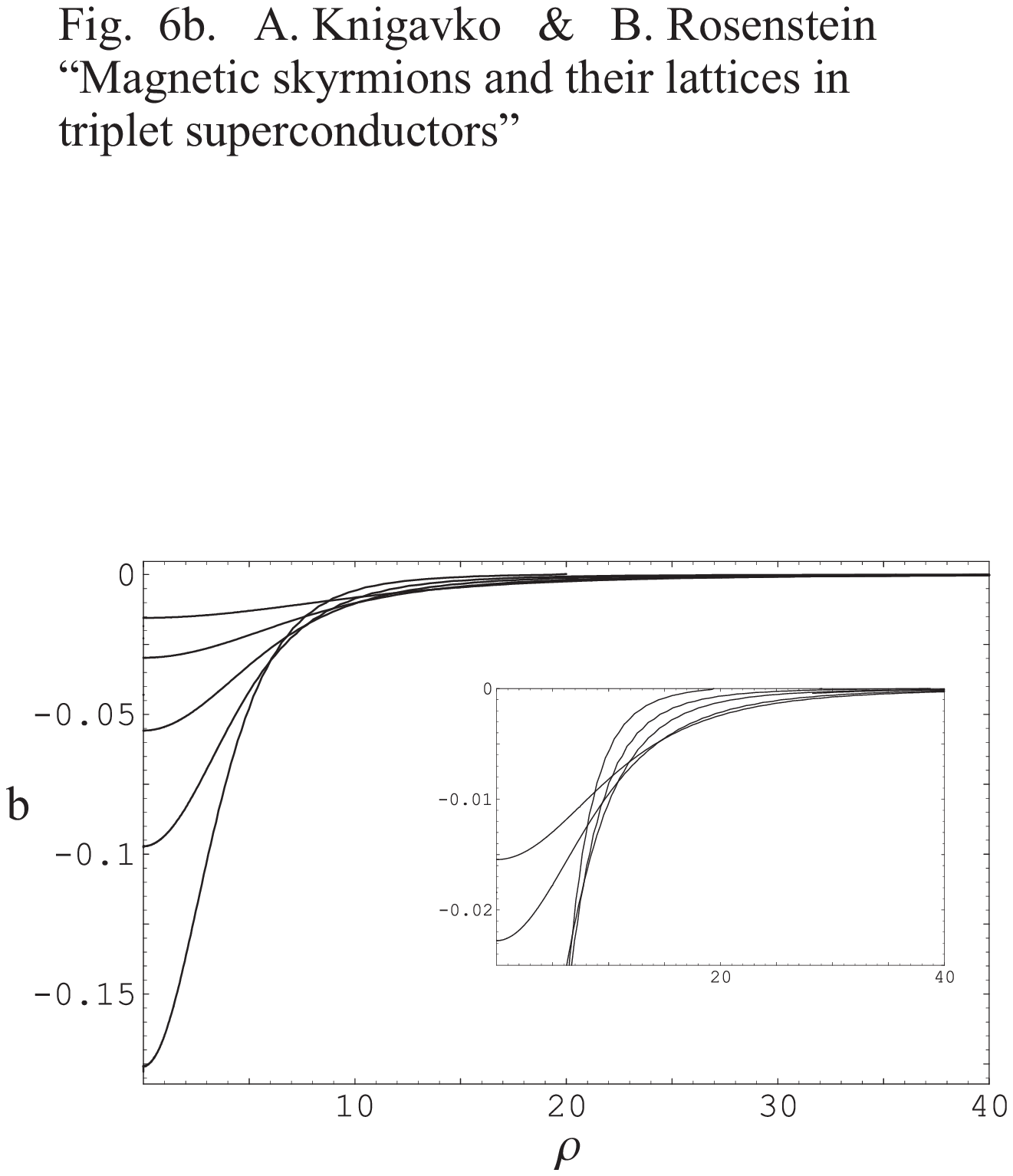,height=7in,width=5.0in,angle=-0}
\end{figure}

\eject
\begin{figure}[htp]
\epsfig{figure=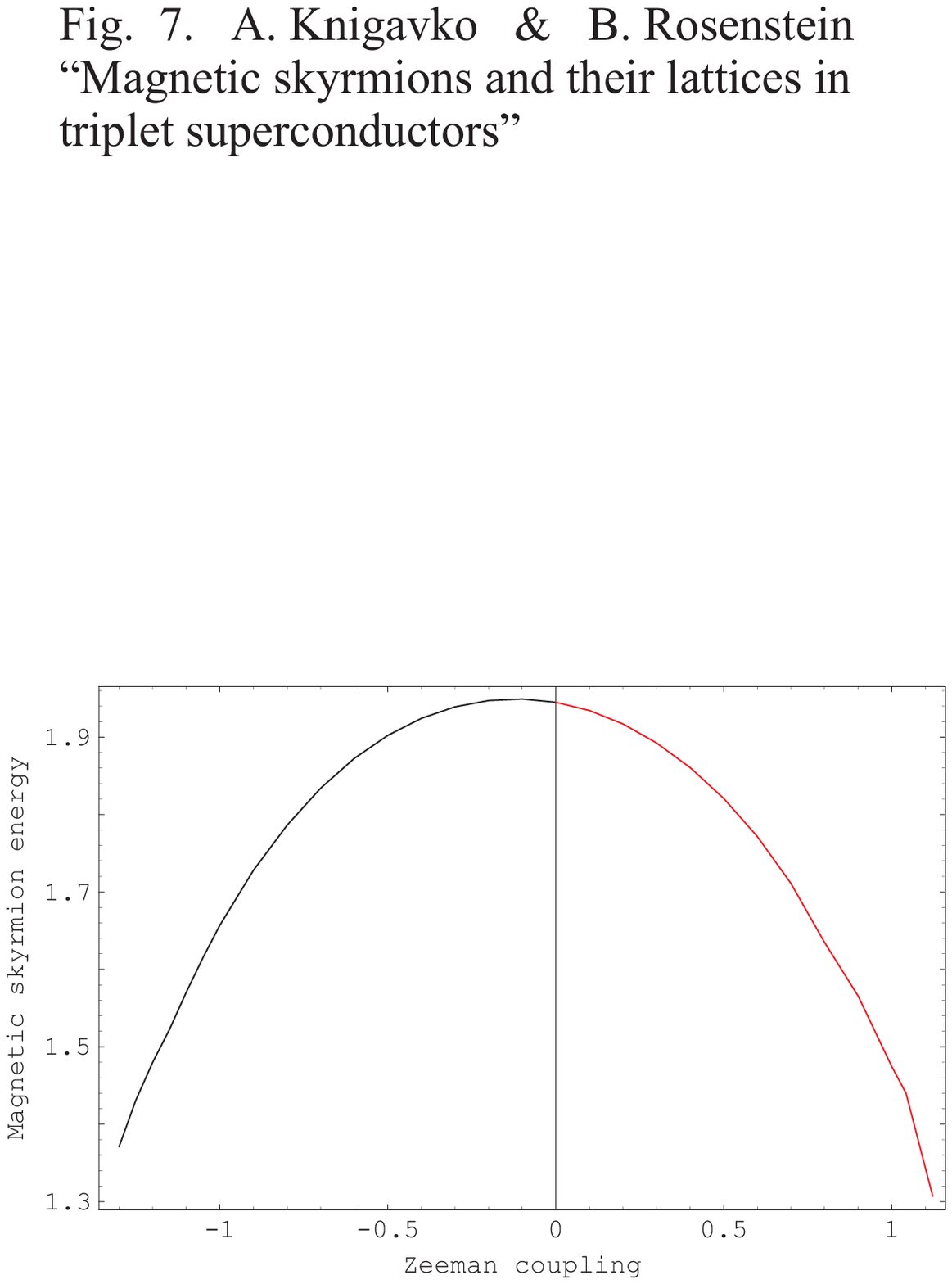,height=7in,width=5.0in,angle=-0}
\end{figure}

\eject
\begin{figure}[htp]
\epsfig{figure=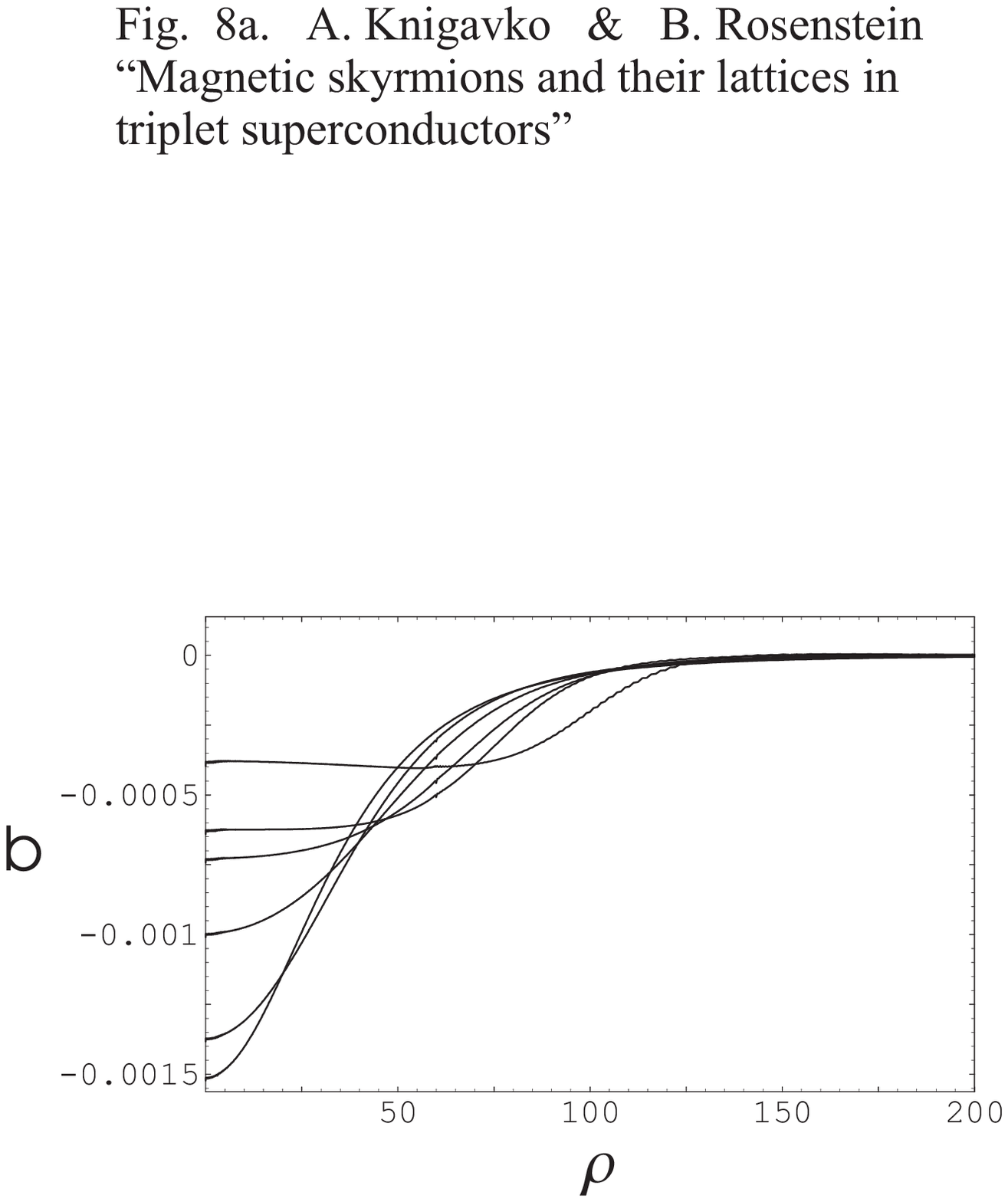,height=7in,width=5.0in,angle=-0}
\end{figure}

\eject
\begin{figure}[htp]
\epsfig{figure=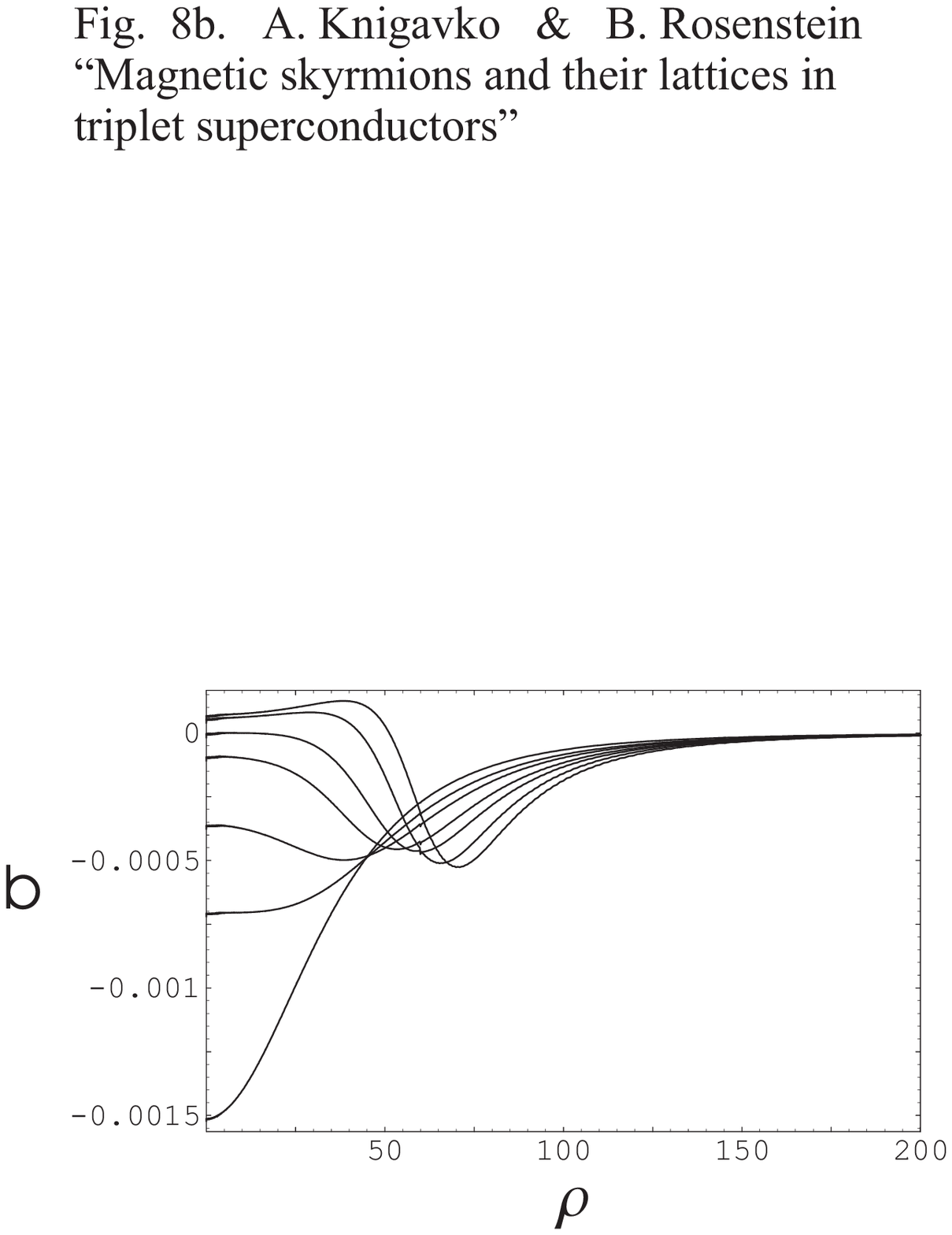,height=7in,width=5.0in,angle=-0}
\end{figure}

\end{document}